\documentstyle{article}

\newskip\humongous \humongous=0pt plus 1000pt minus 1000pt

\newif\ifdtup

\def\oldreffmt#1{\rlap{[#1]} \hbox to 2\parindent{}}

\def\figfmt#1{\rlap{Figure {#1}} \hbox to 1in{}}

\def\beq{\begin{equation}}
\def\eeq{\end{equation}}

\font\tenbf=cmbx10
\font\tenrm=cmr10
\font\tenit=cmti10
\font\elevenbf=cmbx10 scaled\magstep 1
\font\elevenrm=cmr10 scaled\magstep 1
\font\elevenit=cmti10 scaled\magstep 1

\font\ninerm=cmr9

\def\VT{ V(\phi,T)}
\def\V{V(\phi)}
\def\VTI{ V^I(\phi,T)}
\def\excite#1{${}^{#1}$}

\renewenvironment{thebibliography}[1]
 { \elevenrm
   \begin{list}{\arabic{enumi}.}
    {\usecounter{enumi} \setlength{\parsep}{0pt}
     \setlength{\itemsep}{3pt} \settowidth{\labelwidth}{#1.}
     \sloppy
    }}{\end{list}}
\begin{document}
\begin{center}{{\tenbf ASPECTS OF THE\\
               \vglue 10pt
               ELECTROWEAK PHASE TRANSITION\footnote{Presented at
               the 1992 Meeting of the Division of Particles \& Fields
               of the APS (DPF 92), Batavia, IL, November 10-14, 1992.}\\}
\vglue 1.0cm
{\tenrm Patrick Huet \\}
\baselineskip=13pt
{\tenit Stanford Linear Accelerator Center\\}
\baselineskip=12pt
{\tenit Stanford University, Stanford, CA 94309\\}
%\vglue 0.3cm
%{\tenrm and\\}
\vglue 0.8cm
{\tenrm ABSTRACT}}
\end{center}
\vglue 0.3cm
{\rightskip=3pc
 \leftskip=3pc
 \tenrm\baselineskip=12pt
 \noindent
   The electroweak phase transition is reviewed in light of some recent
developments. Emphasis is on the issue whether the transition is first
or second order and its possible role in the generation of the baryon
asymmetry of the universe.
%\vfill
%\begin{flushleft}
%Parallel Session Suggestion: {J. Particle Astrophysics(T)}\\
%Title: {Aspects of the Electroweak Phase Transition}\\
%Presenting Author: {Patrick Huet}\\
%Postal Address: {SLAC  P.O. Box 4349, STANFORD CA 94309}\\
%E-mail Address: {Huet@slacvm}\\
%Telephone: {415-926-4434}\\
%FAX: {415-926-4500}\\
%\end{flushleft}
\vglue 0.6cm}
{\elevenbf\noindent 1. Introduction}
\vglue 0.4cm
\baselineskip=14pt
\elevenrm
Key features of the observable universe are often thought to be
the results of
transient phenomena which occurred in the course of its evolution. The
electroweak phase transition (EWPT) is such a phenomenon. However,
until recently, it was thought to bear no consequences for today's
universe. In a classic paper, Kuzmin, Rubakov and Shaposhnikov showed
otherwise.\excite{1} They uncovered the possibility that the matter
antimatter asymmetry of today's world could have been produced during
the EWPT through non-perturbative physics. In their pioneering work,
they established the necessity of a first order phase transition.
Various groups have since proposed a wealth of explicit
mechanisms.\excite{2,3,4} It has become imperative to broaden our
understanding of the EWPT in order to transform a scenario of
baryogenesis into an actual prediction of the baryon asymmetry of the
universe (BAU) that can be confronted with observations. In the following,
I will describe recent progress made in this direction using the
minimal standard model as a prototype.
\vglue 0.5cm
{\elevenbf\noindent 2. First Order vs. Second Order}
\vglue 0.2cm
{\elevenit \noindent 2.1. The Effective Potential}
\vglue 0.1cm
The standard method of determining the order of the EWPT is to compute
the effective potential $\VT$ for the higgs vev $\phi$, taking into
account the coupling of the vacuum to a thermal bath of particles at a
temperature of about $100$ GeV. The calculations are usually done in
the imaginary time formalism. Here is a real time picture. Let us
construct a fictitious $\phi$-wall interpolating between the two phases
and held steady in the plasma by adequate sources. The goal is to
compute $\VT$ across the wall, that is, minus the total pressure in the
plasma. The unbroken phase is filled with a gas of relativistic
particles whose pressure is known to be $g^* {\pi^2 \over 90}T^4$ with
$ g^* \sim 100$. The pressure in a given region of space is the latter
supplemented with a pure higgs contribution $-\V$ and the total
momentum exchanged between the wall and the plasma integrated up to
this point. The latter is easily computed for a single
particle via conservation of energy, $ {\bf k}^2+ m(\phi)^2 =$
constant.\excite{5} This yields the following result
\beq \VT = \V - g^* {\pi^2
\over 90}T^4 - \sum_{species} \int_0^{m(\phi)^2} dm^2 \int{d^3k \over
(2 \pi)^3 } { n(E) \over 2E} + \bullet \bullet \bullet
\eeq
\beq \VT =
D(T^2-T_o^2) \phi^2 - E T \phi^3 + {\lambda_T \over 4} \phi^4 + \bullet
\bullet \bullet
\eeq
with $D$, $E$, $\lambda_T$ and $T_o$ functions of the couplings of the
theory. The cubic term  originates from the bosonic sector: $E \sim
m_W^3,\, m_Z^3$. To see that, note that the Bose-Einstein distribution
$n(E)$ behaves as ${T \over m}$, as ${\bf k}\rightarrow 0$. Inputting
this information in Eq.~(1) readily generates a term $\sim m^3$. The
dots indicate that Eq.~(1) corresponds to a one loop calculation. It is
known that there are infrared divergences at higher order in the
bosonic sector which contribute effectively as $({ gT \over m})^n$. One
can worry whether, once properly taken into account, they wouldn't wash
away the one loop infrared effect. An answer, using imaginary time
techniques, was recently proposed.\excite{6,7} Its real time
counterpart goes as follows. Gauge interactions in the plasma affect the low
momentum behavior of the gauge bosons, giving them an additional mass
$\Pi$ which is non-vanishing in the limit of small $\phi$. $\Pi$ is the
polarization tensor of the zero modes of the gauge bosons computed in
{\elevenit imaginary time}. This amounts to reducing the population of
the long wavelength excitations of the gauge fields from $T/m$ to the
smaller value $T/\sqrt{m^2+\Pi}$ by screening them with a cloud of
$SU(2)$ charged particles; a dramatic effect in the limit $\phi
\rightarrow 0$. An infrared improved effective potential
$\VTI$ can then be computed by the above method with the replacement
$n(E) \rightarrow n^{\Pi}(E) = ($exp$ \beta\sqrt{{\bf k}^2 + m^2 + \Pi}
-1)^{-1} $. One obtains Eq.~(2) with the substitution
\beq
 -ET\phi^3 \rightarrow - {T \over 12 \pi}
\sum_{W_l,Z_l,W_\bot, Z_\bot}( m^2(\phi)+\Pi)^{3/2} -
(\Pi)^{3/2}
\eeq
$\Pi_{W_l,Z_l}$ can be computed perturbatively. In leading order, it is
$\sim g^2T^2$ and dominates the mass term in the interpolating wall
between the two phases; in such a case, its contribution to Eq.~(3) is
polynomial in $\phi^2$ which only slightly corrects the parameters $D$
and $\lambda_T$. $\Pi_{W_\bot,Z_\bot}$ is not computable perturbatively
but is believed\excite{8} to be at most $\sim g^4T^2$; in such a case,
it can be ignored\footnote{\ninerm \baselineskip=11pt
Except in a region of small $\phi$ which
doesn't affect the qualitative behavior of $\VTI$.\excite{9}} and the
RHS of Eq.~(3) turns into a term cubic in $\phi$. As a result, one obtains
\beq
\VTI = D^*(T^2-T_o^2)
\phi^2 -{\bf 2\over3} E T \phi^3 + {\lambda^*_T \over 4} \phi^4 +
\ldots
\eeq
where now the dots refer to perturbative corrections. The presence of
the cubic term in this improved effective potential implies a first
order phase transition. This conclusion is subject to caveats. (1)~ The
phase transition would be of $2^{nd}$ order if the ``magnetic mass"
$\Pi_\bot$ were found to be significantly larger.
(2) Gleiser and Kolb\excite{10}
argued, on the basis of an $\epsilon$-expansion\excite{17}
applied to a simpler
system with similar small $\phi$ behavior,
that
Eq.(4) doesn't describe properly long range fluctuations in the
scalar sector.\footnote{\ninerm \baselineskip=11pt
These fluctuations are ignored in Eq.(4).
An assumption believed to be reasonable if the higgs mass is below $150$ GeV.}
%They generalize conclusions obtained using an
%$\epsilon$-expansion applied to a similar but simpler system
They expect a more weakly first order phase transition.
%They constructed a model which mimic the behavior of $\VTI$  for small
%$\phi$ to which they applied an $\epsilon$-expansion.  This analysis
%lead them to suspect a more weakly first order phase transition.
\vglue 0.2cm
{\elevenit \noindent 2.2. Completion of the Phase Transition}
\vglue 0.1cm
The universe supercools until thermal fluctuations are able to
destabilize the system at a significant rate. This corresponds to the
nucleation of ``critical bubbles". These bubbles evolve to a
macroscopic size $ \sim 10^{-5} {M_{planck} \over T^2}$ before they
collide and fill the universe. During this short period (${\delta T
\over T}\sim 10^{-7}$), baryogenesis takes place in the
propagating bubble walls. Alternatives to the scenario above have been
proposed. For instance, Kolb and Gleiser\excite{10} have argued that
long range fluctuations are so rapid at the phase transition that the
universe is more adequately described by an emulsion of subcritical
domains of both vacua which smoothly interpolates between the two
phases as the universe cools down. This scenario has been strongly
criticized\excite{11} and shown to be possibly relevant for a range
of parameters orthogonal to the ones which allow
baryogenesis. This comes about by requiring the
freezing-out of the baryon violating processes in the broken phase in
order to prevent the washing-out of the BAU; in the standard minimal
model\excite{7} and with the use of Eq.~(4), this requires a
higgs mass no larger than $40$ GeV, far below the experimental limit.
\vglue 0.5cm
{\elevenbf \noindent 3. Bubble Wall Dynamics \hfil}
\vglue 0.2cm
{\elevenit \noindent 3.1. Bubble Wall Velocity} \vglue 0.1cm All the
scenarios of baryogenesis make convenient assumptions on the shape and
velocity of the wall. In the scenarios making use of the quantum
mechanical reflection of top quarks,\excite{4} the wall
thickness is assumed to be of the order of the Compton
wavelength of the reflected particles in order to prevent an excessive
suppression. This condition requires particular conditions: the wall
thickness being typically one or two orders of magnitude too large. In
the scenarios of baryogenesis inside the wall,\excite{3} the
velocity is assumed to be large enough to prevent the BAU to diffuse
and be washed out in the unbroken phase but it is assumed to be slow
enough to maximize the production rate.\excite{12,3} The physics of
the damping of the wall was only recently understood.\excite{12,7}
The moving wall sets the plasma out of equilibrium $n(E)
\rightarrow n(E) + \delta n(E)$ by an amount proportional to the
velocity. This excess $\delta n(E)$ generates, in turn, according to
Eq.~(1) an additional component to the pressure which grows until it
balances the difference in pressure across the wall. This condition
fixes the velocity $\gamma v$ to be in the range of $0.05$ to 1,
depending on the parameters. A range favorable for baryogenesis.
\vglue 0.2cm
{\elevenit \noindent 3.2. Bubble Wall Stability}
\vglue 0.1cm
A velocity smaller than the speed of sound $\sim 0.6$ is characteristic
of a deflagration process. It is common belief that a deflagration
front is unstable under perturbations whose size is large enough to
overcome the surface tension $\sim {\sigma \over \lambda^2}$. These
instabilities have been contemplated in the context of both the
EWPT\excite{14} and the QCD phase transition.\excite{15} However,
Landau's original stability analysis\excite{13} was designed for
violent macroscopic phenomena. A recent linear stability
analysis\excite{16} was tailored for more general phenomena and in
particular, for the EWPT where it was shown that no perturbation can
destabilize the moving wall in the allowed range of velocity. The
reason goes as follows. A perturbation of the front triggers
fluctuations in the temperature and velocity of the plasma in both
phases; all of which are entangled by conservation of energy-momentum.
Additional information on the microscopic dynamics has to be input to
determine completely the subsequent evolution of the perturbation.
Landau assumed that the relative velocity wall-plasma is unaffected by
the fluctuations.  However, we learned above that, in the EW case, the
wall-plasma velocity is proportional to $\VTI$, a sensitive function of
the temperature, and, consequently, varying significantly as the
temperature fluctuates. This effect tends to oppose to the growth of
the perturbation. This sensitivity is measured by a dimensionless
parameter which turns out to be so large $ \sim {1 \over
\alpha_W}$ that it prevents the perturbation from growing at all.
\vglue 0.4cm
{\elevenbf\noindent 5. References \hfil}
\vglue 0.4cm

\end{document}